# Towards Critical Clearing Time Sensitivity for DAE Systems with Singularity

Chetan Mishra and Chen Wang
Dominion Energy
Richmond, VA 23220, USA

Xin Xu
Department of EECS
University of Tennessee
Knoxville, TN 37996, USA

Virgilio A. Centeno
Department of ECE
Virginia Tech
Blacksburg, VA 24060, USA

*Abstract*—Standard power system models are parameter dependent differential-algebraic equation (DAE) type. Following a transient event, voltage collapse can occur as a bifurcation of the transient load flow solutions which is marked by the system trajectory reaching a singular surface in state space where the voltage causality is lost. If a fault is expected to cause voltage collapse, preventive control decisions such as changes in AVR settings need to be taken in order to get enhance the system stability. In this regard, the knowledge of sensitivity of critical clearing time (CCT) to controllable system parameters can be of great help. The quasi-stability boundary of DAE systems is more complicated than ODE systems where in addition to unstable equilibrium points (UEP) and periodic orbits, singularity plays an important role making the problem challenging. The stability boundary is then made up of a number of dynamically distinct components. In the present work, we derive the expression for CCT sensitivity for the phenomenon where the critical fault-on trajectory intersects the singular surface itself which is one such component forming the stability boundary. The results are illustrated for a small test system in order to gain visual insights.

## I. INTRODUCTION

Utilities tend to maximize the utilization of the existing transmission network in certain regions due to difficulties in building new right-of ways to supply the increasing demand. This coupled with loss in voltage controllability due to retiring conventional generation has made voltage instability a serious concern.

In the past, voltage collapse was studied as a small signal problem [1] resulting from the saddle node bifurcation (SNB) of load flow solutions where a stable equilibrium point (SEP) merges with a UEP on its stability boundary and vanishes. However, during transient conditions, voltage collapse can occur in a different manner [2]. A qualitative change to the system that is highly conducive to this phenomenon is singularity induced bifurcation [3] where one or more equilibrium points (EP) merge with the singular surface of the algebraic constraint. On the singular surface, algebraic variables like load bus voltage lose the causal relationship with dynamic states like generator rotor angle which has been shown to have a strong relationship with voltage collapse [Hiskens and Hill [4]. Trajectories passing through the singular surface may bifurcate and settle to an infeasible (low voltage) point. However, the DAE model cannot predict the dynamics on the singular surface [5] thus requiring modeling of load dynamics. Nevertheless, singularity/loss of voltage causality serves an important purpose as a precursor for voltage collapse.

CCT refers to the maximum time that can be taken to clear the fault while remaining stable which is popularly used as a metric for stability margin. There has been plenty of work in the past on CCT sensitivity computation. Ayasun [6] reduced the multimachine system to single machine infinite bus system to evaluate sensitivities which is computationally efficient yet approximate. Nguyen [7] and Laufenberg [8] computed sensitivity of angle and speed trajectory in the post fault phase w.r.t. fault clearing time which are expected to grow for marginally stable trajectories. Nguyen also computed CCT sensitivities by approximating the relevant portion of stability boundary by constant energy surface passing through the controlling unstable equilibrium point (CUEP). One of the more recent works by Dobson et.al. [9] does not make this approximation for stability boundary and simply uses a local characterization of it to give more accurate estimates. His derivation is for unconstrained ODE systems and an extension is proposed for DAE systems under the assumption that the voltage causal region totally contains the stability region (SR) of the SEP of interest within the range of parameter changes. Our recent work [10] presented the derivations for CCT sensitivity expressions for ODE type systems with inequality constraints. Since the definition of CCT for DAE systems has an added constraint of not reaching singularity (voltage collapse), it is extremely important to incorporate that when deriving the expression for CCT sensitivity which will be the focus of this work.

In Section II. , The DAE model is described along with the stability theory of such systems taking into account the role of singular surface. The derivation of CCT sensitivity expression for one out of three phenomena of loss of stability through singularity is presented in Section III. Finally, the derivations are validated through simulation on a one bus one machine system in Section IV.

## II. SYSTEM MODEL AND STABILITY THEORY OF DAE SYSTEMS

The system being considered in this work is defined by the following state equation,
$$\dot{x} = f(x, y) \qquad (1)$$
$$0 = g(x, y)$$

Here, $x \in R^n$ are dynamic states such as generator rotor angles, generator flux linkages, etc. and $y \in R^m$ are algebraic states such as load bus voltages and phase angles. The equality constraint $g = 0$ which is given for each configuration i.e. pre-fault, fault-on and post-fault, gives the corresponding surface in the overall state space on which the system evolves. The system jumps between these surfaces whenever switching happens with $x$ varying smoothly in time (given by the first equation in (1)) whereas $y$ jumps during switching. On a given surface, as long as $\frac{\partial g}{\partial y}$ is invertible, the trajectory exists and is one dimensional. Points where there it isn't true are called singular points and are given by,

$$S = \{x, y | g(x, y) = 0, \Delta(x, y) = \det\left(\frac{\partial g}{\partial y}\right) = 0\} \quad (2)$$

Subset of state space which contains the stable equilibrium point (SEP) of interest and where, without loss of generality, all eigen values of $\frac{\partial g}{\partial y}$ are positive is the region of interest and contains the SR. To help characterize critical elements on the stability boundary of DAE systems, a transformed system was proposed in [11] as shown below.

$$\dot{x} = \Delta(x, y) \times f(x, y) \quad (3)$$
$$\dot{y} = -\kappa(x, y) = -adj\left(\frac{\partial g(x, y)}{\partial y}\right) \times \frac{\partial g(x, y)}{\partial x} \times f(x, y)$$

The above system is a time scaled version of the original DAE system which is why their stability region and its boundary are same. A nice quality of this system is that it gets rid of the singularities of the original DAE system. However, we cannot use it directly for our derivations of CCT sensitivity because the concept of time is different from the original system. Additionally, this system introduces new critical elements on the singular surface. The first category is called semi-singular points where the transformed system trajectory grazes (is tangential to) the singular surface (boundary between shaded and unshaded regions) as shown in Figure 1. These can be characterized as,

$$Semi-Singular: \Xi = \left\{(x, y) \in S \middle| \frac{\partial \Delta(x, y)}{\partial y} \times \kappa(x, y) = 0\right\} \quad (4)$$

For the characterization of quasi-stability boundary ($n - 1$ dimensional) which is more relevant from engineering viewpoint, of particular importance are $n - 2$ dimensional connected components in $\Xi$.

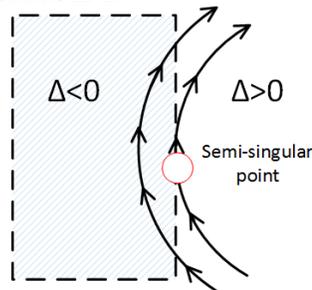

**Figure 1 Dynamics Near Semi-Singular**

The second category of points are called pseudo equilibrium points which are EPs of the transformed system but not of original system. These are defined as,

$$Pseudo\ EP: \Psi = \{x, y \in S | \kappa(x, y) = 0\} \quad (5)$$

Of particular importance are $n - 2$ dimensional connected components of pseudo EPs. These have $n - 2$ dimensional center manifold which is the connected component itself and 2 non-zero eigen values. Depending on the sign of those eigen values, the points can be characterized as source (both positive), sink (both negative) or saddle (one positive one negative). Saddle type points are the crucial for characterizing the stability boundary of the DAE system. The dynamics in its vicinity are shown in Figure 2 where the arrows point in the direction of the flow.

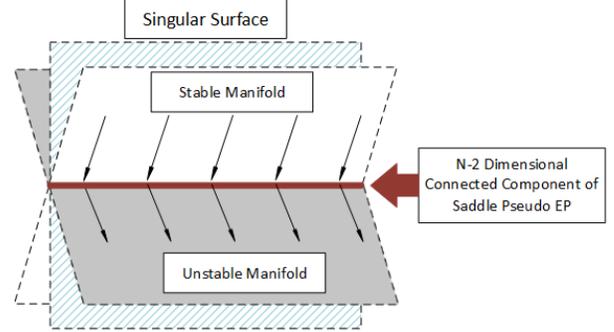

**Figure 2 Dynamics near Saddle Pseudo EP**

Under reasonable assumptions [11], the quasi-stability boundary of the transformed system and therefore the original DAE system is comprised of –
1. $n - 1$ dimensional components of singular surface
2. Stable manifold of $n - 2$ dimensional connected components of semi-singular points
3. Attracting set of $n - 2$ dimensional saddle type pseudo EPs.
4. Stable manifolds of type-1 UEP and periodic orbits

### III. CCT SENSITIVITY DERIVATION

#### A. Overview

A critical trajectory for a given fault for a given value of system parameter $p$ would be one in which the fault is cleared at its CCT. Let the base parameter value be denoted by $p^*$ with the term base critical trajectory referring to the critical trajectory for $p = p^*$. Also, by definition, the state variable value at CCT must lie on one of the above components in state space. Under $p$ variations, the stability boundary will change and so will the system trajectory. For the new fault-on trajectory to intersect the new stability boundary, the fault clearing time will have to be adjusted. The amount of adjustment required per unit change in $p$ gives the CCT sensitivity. The overall process to CCT sensitivity computation around the base critical trajectory involves the following steps –
1. Find the sensitivity of the state variable value at the fault clearing time with respect to fault clearing time and parameter evaluated at the CCT of the base critical trajectory.
2. Find the sensitivity of the stability boundary to parameter changes evaluated at state variable value at the CCT of base critical trajectory.
3. Equate the above two to get CCT sensitivity.

Now, we don't need to compute the sensitivity of the whole stability boundary but just the component relevant [12] to the particular fault. Furthermore, the equation of the stability boundary is hard to find and usually only a local approximation around a point lying on it is available. For example, the equation of stable manifold of type-1 UEP is locally approximated near the UEP by a hyperplane normal to the unstable eigen vector. If the local approximation is not given at the exit point (where fault on trajectory intersects the stability boundary component) and therefore is given at some point along the post-fault trajectory (more common), the above steps will need to be evaluated at that point and not the exit point. In that case, in the first step, the sensitivity of state variable value is to be found w.r.t. fault clearing time, $p$ and the time spent along the post-fault trajectory.

In [9] and [10], the derivations have been done for the last component (stable manifold of type-1 UEP). In the present paper, we will be presenting the derivation for the first component which is when the base critical fault-on trajectory intersects the singular surface i.e. the singular surface is the relevant component of the stability boundary. Therefore, the CCT in this case is synonymous to how long it takes for the fault-on trajectory to reach direct voltage collapse/singularity. There is a greater value to understanding the sensitivity of time to reach singular surface than being a CCT sensitivity. This number can be computed for any fault regardless of what the phenomenon of instability is as it will give an insight into what control parameters can efficiently *push away* the singular surface thereby reducing the likelihood of voltage collapse. As a note, derivations are done assuming $p$ as scalar.

### B. Sensitivity of the State Value at Fault Clearing

In a typical study scenario for TSA of DAE systems, we have pre-fault, fault-on and post-fault constraint surfaces given by their respective algebraic constraints $g = 0$ with the system jumping between those as switching happens. Let us denote the value of $y$ on fault-on constraint surface for a given value of $x$ as $y_{fault}$ i.e. $g_{fault}(x, y_{fault}, p) = 0$. We can similarly define $y_{post}$ which is the value of $y$ immediately after the fault is cleared so $g_{post}(x, y_{post}, p) = 0$ and $y_{pre}$ for $y$ value if the fault were to immediately clear by reverting back to the pre-fault system conditions(topology, etc).

In TSA, its usually assumed that the system operating point at $t = 0$ denoted by $(x^0, y_{pre}^0)$ is the SEP of pre-fault system,
$$f_{pre}(x^0, y_{pre}^0, p) = 0 \tag{6}$$
$$g_{pre}(x^0, y_{pre}^0, p) = 0$$

Thus, $x^0$ lies on a one-dimensional manifold making it locally a function of $p$. The sensitivity of the $x^0$ to $p$ evaluated at base critical trajectory's pre-fault SEP and $p^*$ is then given as,
$$\left.\frac{\Delta x^0(p)}{\Delta p}\right|_{p^*} = A_1^{(n \times 1)} = \left[\frac{\partial f_{pre}}{\partial x} - \frac{\partial f_{pre}}{\partial y} \times \left[\frac{\partial g_{pre}}{\partial y}\right]^- \times \frac{\partial g_{pre}}{\partial x}\right]^- \tag{7}$$
$$\times \left(\frac{\partial f_{pre}}{\partial y} \times \left[\frac{\partial g_{pre}}{\partial y}\right]^- \times \frac{\partial g_{pre}}{\partial p}\right.$$
$$\left. - \frac{\partial f_{pre}}{\partial p}\right)\bigg|_{x_{pre}^{s*}, y_{pre}^{s*}, p^*}$$

Here, SEP of base pre-fault system is denoted by $x^{s*}, y_{pre}^{s*}$ i.e. $f_{pre}(x^{s*}, y_{pre}^{s*}, p^*) = g_{pre}(x^{s*}, y_{pre}^{s*}, p^*) = 0$. The corresponding value of $y_{fault}$ f is denoted by $y_{fault}^{0*}$.

Let $\left(\varphi_{fault}^x(x^0, y_{fault}^0, t, p), \varphi_{fault}^{y_{fault}}(x^0, y_{fault}^0, t, p)\right)$ represent the generalized parametric flow of $(x, y)$ for the fault-on DAE system starting from any point $(x^0, y_{fault}^0)$ i.e. $\frac{\partial \varphi_{fault}^x}{\partial t} = f_{fault}\left(\varphi_{fault}^x, \varphi_{fault}^{y_{fault}}, p\right), g_{fault}\left(\varphi_{fault}^x, \varphi_{fault}^{y_{fault}}, p\right) = 0$, $\varphi_{fault}^x(x^0, y_{fault}^0, 0, p) = x^0$, $\varphi_{fault}^{y_{fault}}(x^0, y_{fault}^0, 0, p) = y_{fault}^0$.

Here, since $(x^0, y_{fault}^0)$ is assumed to not be on the singular surface of the fault-on system, by implicit function theorem, $y_{fault}^0$ can locally be written as a function of $x^0$ and therefore $\varphi_{fault}^x, \varphi_{fault}^{y_{fault}}$ can be written purely as a function of $x^0$ and $p$.

Given $x^0$ lies on the SEP of pre-fault system, let the value of $x$ at any fault clearing time $t^{cl}$ be denoted by $x^{cl} = \varphi_{fault}^x(x^0(p), t^{cl}, p)$. Next, the sensitivity of $x^{cl}$ is evaluated at the fault clearing time of base critical trajectory which is also its CCT by definition and is denoted by $t^{cr*}$. Let state value of base critical fault on trajectory at the time of fault clearing be denoted by $\left(x^{cl*}, y_{fault}^{cl*}\right)$ i.e. $\left(x^{cl*}, y_{fault}^{cl*}\right) = (\varphi_{fault}^x(x_{pre}^{s*}, t^{cr*}, p^*), \varphi_{fault}^{y_{fault}}(x_{pre}^{s*}, t^{cr*}, p^*))$. The sensitivity can then be computed as,
$$\left.\frac{\Delta x^{cl}}{\Delta p}\right|_{p^*} = B_1 \times \left.\frac{\Delta x^0}{\Delta p}\right|_{p^*} + B_2 \times \left.\frac{\Delta t^{cl}}{\Delta p}\right|_{p^*} + B_3 \tag{8}$$

Where,
$B_1^{(n \times n)} = \left.\frac{\partial \varphi_{fault}^x}{\partial x^0}\right|_{t^{cl*}, p^*}, B_2^{(n \times 1)} = \left.\frac{\partial \varphi_{fault}^x}{\partial t}\right|_{t^{cl*}, p^*} = f_{fault}(x, y_{fault}, p)\big|_{x^{cl*}, y_{fault}^{cl*}}, B_3^{(n \times 1)} = \left.\frac{\partial \varphi_{fault}^x}{\partial p}\right|_{t^{cl*}, p^*}$

Finally, substituting the expression from Eqn (7) into Eqn (8),
$$\left.\frac{\Delta x^{cl}}{\Delta p}\right|_{p^*} = B_1 \times A_1 + B_2 \times \left.\frac{\Delta t^{cl}}{\Delta p}\right|_{p^*} + B_3 \tag{9}$$

### C. Notes on Trajectory Sensitivity of DAE with Singularities

$B_1$ and $B_3$ in equation (8) are computed through trajectory sensitivity analysis [13] of the fault on system using the following variational equation for a generic parameter $param$ which can be $x^0$ and/or $p$,
$$\frac{\partial \varphi_{fault}^{\dot{x}}}{\partial param} = \frac{\partial f_{fault}(x, y_{fault}, p)}{\partial x} \times \frac{\partial \varphi_{fault}^x}{\partial param} + \frac{\partial f_{fault}}{\partial y_{fault}} \times \frac{\partial \varphi_{fault}^{y_{fault}}}{\partial param} \tag{10}$$
$$+ \frac{\partial f_{fault}}{\partial param}$$
$$0 = \frac{\partial g_{fault}(x, y_{fault}, p)}{\partial x} \times \frac{\partial \varphi_{fault}^x}{\partial param}$$
$$+ \frac{\partial g_{fault}(x, y_{fault}, p)}{\partial y_{fault}} \times \frac{\partial \varphi_{fault}^{y_{fault}}}{\partial param}$$
$$+ \frac{\partial g_{fault}(x, y_{fault}, p)}{\partial param}$$
$$0 = \frac{\partial g_{post}(x, y_{post}, p)}{\partial x} \times \frac{\partial \varphi_{fault}^x}{\partial param} + \frac{\partial g_{post}(x, y_{post}, p)}{\partial y_{post}} \times \frac{\partial \varphi_{fault}^{y_{post}}}{\partial param}$$
$$+ \frac{\partial g_{post}(x, y_{post}, p)}{\partial param}$$

Here, $\varphi_{fault}^{y_{post}}$ represents the $y$ value on the post-fault surface evaluated along the fault-on trajectory. It can be clearly seen from the third equation above that on the singular surface of the post fault system, $\frac{\partial \varphi_{fault}^{y_{post}}}{\partial param}$ blows to infinity which introduces a big challenge to CCT sensitivity computation as will be seen later.

## D. CCT Sensitivity Derivation for Fault-On Trajectory Exiting Through Post-Fault System's Singular Surface

The fault-on system is assumed to not have any singularities within region of interest in state space. Therefore, post-fault system's singularity is our focus in this work. The phenomenon of loss of stability will be one where the fault-on trajectory intersects the post-fault systems's singular surface directly if the fault were to be cleared i.e. $\{\Delta_{post}(x^{cl}, y_{post}^{cl}, p) = 0, g_{post}(x^{cl}, y_{post}^{cl}, p) = 0\}$ which also locally characterizes the stability boundary. Here, $y_{post}^{cl}$ represents the $y$ value immediately following the fault clearing. $y_{post}^{cl*}$ denotes the corresponding value for the base critical trajectory. Therefore, for $t^{cl}$ so serve as a CCT under variation of $p$, $\frac{\Delta x^{cl}}{\Delta p}, \frac{\Delta y_{post}^{cl}}{\Delta p}$ around the base critical trajectory should satisfy,

$$\frac{\partial \Delta_{post}(x,y,p)}{\partial x}\bigg|_{x^{cl*}, y_{post}^{cl*}, p^*} \times \frac{\Delta x^{cl}}{\Delta p}\bigg|_{p^*} \quad (11)$$
$$+ \frac{\partial \Delta_{post}(x,y,p)}{\partial y}\bigg|_{x^{cl*}, y_{post}^{cl*}, p^*} \times \frac{\Delta y_{post}^{cl}}{\Delta p}\bigg|_{p^*}$$
$$+ \frac{\partial \Delta_{post}(x,y,p)}{\partial p}\bigg|_{x^{cl*}, y_{post}^{cl*}, p^*} = 0$$

$$\frac{\partial g_{post}(x,y,p)}{\partial x}\bigg|_{x^{cl*}, y_{post}^{cl*}, p^*} \times \frac{\Delta x^{cl}}{\Delta p}\bigg|_{p^*}$$
$$+ \frac{\partial g_{post}(x,y,p)}{\partial y}\bigg|_{x^{cl*}, y_{post}^{cl*}, p^*} \times \frac{\Delta y_{post}^{cl}}{\Delta p}\bigg|_{p^*}$$
$$+ \frac{\partial g_{post}(x,y,p)}{\partial p}\bigg|_{x^{cl*}, y_{post}^{cl*}, p^*} = 0$$

Now, $\frac{\Delta y_{post}^{cl}}{\Delta p} = \frac{\partial \varphi_{fault}^{y_{fault}}}{\partial param} \times \frac{\Delta param}{\Delta p}$ by definition and as seen before, due to singularity, it will blow to infinity. We propose a small trick to resolve this issue. Since $\frac{\partial g_{post}(x,y,p)}{\partial y}\bigg|_{x^{cl*}, y_{post}^{cl*}, p^*}$ is singular, with a single 0 eigen value, there is a left eigen vector corresponding to that lets call it $v^{*T}$. Pre-multiplying the second equation by it gets rid of the second term resulting in,

$$v^{*T} \times C_1 \times \frac{\Delta x^{cl}}{\Delta p} + v^{*T} \times C_2 = 0 \quad (12)$$

Where $C_1 = \frac{\partial g_{post}(x,y,p)}{\partial x}\bigg|_{x^{cl*}, y_{post}^{cl*}, p^*}$ and $C_2 = \frac{\partial g_{post}(x,y,p)}{\partial p}\bigg|_{x^{cl*}, y_{post}^{cl*}, p^*}$. Substituting equation (9) in the above equation, we get the expression for CCT sensitivity,

$$\boxed{\frac{\Delta t^{cl}}{\Delta p} = -\frac{v^{*T} \times (C_2 + C_1 \times (B_1 \times A_1 + B_3))}{v^{*T} \times C_1 \times B_2}} \quad (13)$$

## E. Overall Computation

Normally, $p$ is a vector. The expressions can be simply be derived independently for each element in $p$ as follows,
1. Time domain simulation (TDS) to find CCT as well as the base critical trajectory for $p = p^*$.
2. If at CCT, singularity of post-fault system (DAE non-convergence) is encountered which is the scenario being addressed in this paper, for each $p_i \in p$ –

i. Evaluate trajectory sensitivity of $x$ of fault on system to $x_{pre}^s$ and $p_i$ using Eqn (9).
ii. Singular Value Decomposition of $U \times \Sigma \times V^T = \frac{\partial g}{\partial y}$ at end point of the critical fault on trajectory. Column of $U$ corresponding to 0 singular value gives us $v^*$ to be used in Eqn (13).

The only major computation is trajectory sensitivity which can easily be parallelized across all $p_i$'s. Therefore, the proposed work is easily scalable to large scale systems.

## IV. RESULTS

In this section, in order to be able to visualize the state space, we use a one bus one machine model [14]. Taking bus voltage angle as a reference, the overall dynamics can be written as follows.

$$f(x,y) = \begin{bmatrix} x_2 + \frac{P_m - \frac{E \times y}{X}\sin(x_1)}{D_l} \\ \frac{(P_m - \frac{E \times y}{X}\sin(x_1) - D_g \times x_2)}{M} \end{bmatrix} \quad (14)$$

$$g(x,y) = \frac{E \times y}{X}\cos(x_1) - \frac{y^2}{X} - Q_l$$

Here, $x_1$ denotes the deviation of generator rotor angle from bus phase angle, $x_2$ denotes generator angular speed deviation, $M$ is the generator inertia constant, $P_m$ is mechanical power input to the generator and also the load at bus (lossless system), $D_g$ is generator damping, $E$ is internal emf of the generator, $y$ is the bus voltage magnitude, $Q_l$ is the reactive power load at the bus, $D_l$ is the load damping factor and $X$ is the total impedance (internal impedance of generator plus transmission line impedance). Singular surface is given by,

$$\{\frac{E}{X}\cos(x_1) - \frac{2y}{X} = 0, -\frac{y^2}{X} + \frac{E \times y}{X}\cos(x_1) - Q_l = 0\} \quad (15)$$

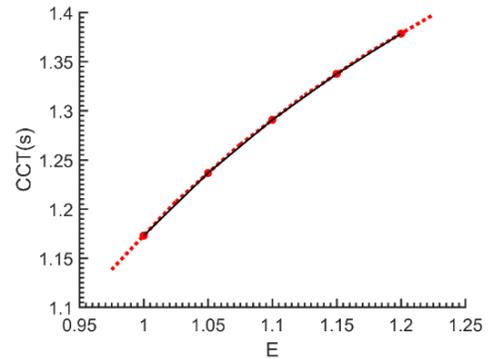

**Figure 3 CCT vs E**

Fault being analyzed is a 3 phase to ground fault on the bus i.e. $g_{fault}(x, y, p) = y = 0$ and cleared without changing the topology (pre-fault and post-fault systems are same). Since the phenomenon of interest is fault on trajectory directly intersecting the post-fault system's singular surface, CCT in this case would be the minimum time the fault needs to be sustained to reach singularity/voltage collapse. The initial value of various parameters are, $X = 0.5, Pm = 0.5, E = 1, M = 1, D_l = 1, D_g = 1, Q_l = 0.1$.

Let us first study the effect of increasing the generator excitation i.e. $p = E$ on CCT which will help understand how much the excitation can help with preventing voltage collapse. The figure above shows the actual CCT vs $E$ obtained through TDS. Also shown by dotted lines are CCT estimates at each marked point using sensitivity formula derived in Section III. It can be seen that the dotted lines are tangent to the original curve which shows the validity of the formula. Also, the trend is as expected where increasing generator excitation helps with post-fault voltage recovery and therefore reduces the changes of voltage collapse.

To further analyze the impact of increasing $E$ visually, we plot the post-fault algebraic constraint surface ($g(x,y,p) = 0$) for different values of $E$ studied (red to pink) along with the post-fault SEP (dot) and the projection of fault on trajectories on those surfaces as shown in Figure 4. The singular surface can be seen as the nose of each surface which is one dimensional. The fault can clearly be seen driving the system towards singularity. Also, for lower values of $E$, the surface is steeper resulting in a more rapid decline in bus voltage and shorter time to singularity.

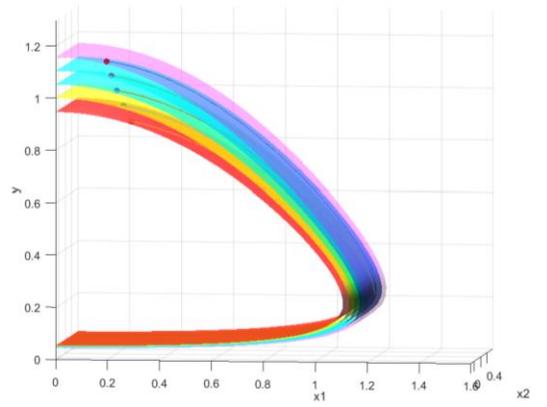

**Figure 4 Constraint Surface vs E**

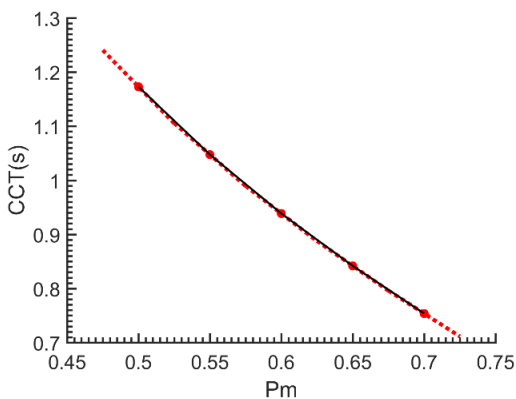

**Figure 5 CCT vs $P_m$**

Next, we validate our results under variation of generator dispatch and correspondingly the real power load. As expected, CCT decreases with increasing generation loading due to the SEP of the post-fault system moving closer to the singular surface. Once again, our derived linear approximation (from CCT sensitivity) is tangential to the actual curve as shown in Figure 5 thus validating our derivation. Furthermore, the acceleration of generator during fault being studied is higher for higher generator loading which further reduces the CCT.

## V. Conclusion and Future Work

There are multiple instability mechanisms for DAE systems with the trajectory reaching a region in state space marked by singularity of algebraic constraints being a characteristic one which is closely related to voltage collapse. CCT sensitivity formula derivation depends on the phenomenon of instability relevant to the fault under study. In the present work, we have focused on derivation for one of the more extreme ones which is observed as a voltage collapse along the fault-on trajectory. A good application for the derived expressions would be in choosing effective preventive controls by pushing away the singular surface thus making it harder to reach. In the future work, we will be deriving CCT sensitivity expressions for the remaining instability phenomena which both result in voltage collapse after some time and not immediately on clearing the fault.

## VI. References

[1] T. van Cutsem and C. Vournas, *Voltage Stability of Electric Power Systems*. Springer US, 1998.
[2] K. L. Praprost and K. A. Loparo, "An energy function method for determining voltage collapse during a power system transient," *IEEE Trans. Circuits Syst. Fundam. Theory Appl.*, vol. 41, no. 10, pp. 635–651, Oct. 1994.
[3] V. Venkatasubramanian, H. Schattler, and J. Zaborszky, "Analysis of local bifurcation mechanisms in large differential-algebraic systems such as the power system," in *Proceedings of 32nd IEEE Conference on Decision and Control*, 1993, pp. 3727–3733 vol.4.
[4] I. A. Hiskens and D. J. Hill, "Failure modes of a collapsing power system," in *Proceedings NSF/ECC Workshop on Bulk Power System Voltage Phenomena II*, 1991, pp. 53–63.
[5] I. Dobson, H.- Chiang, J. S. Thorp, and L. Fekih-Ahmed, "A model of voltage collapse in electric power systems," in *Proceedings of the 27th IEEE Conference on Decision and Control*, 1988, pp. 2104–2109 vol.3.
[6] S. Ayasun, Y. Liang, and C. O. Nwankpa, "A sensitivity approach for computation of the probability density function of critical clearing time and probability of stability in power system transient stability analysis," *Appl. Math. Comput.*, vol. 176, no. 2, pp. 563–576, May 2006.
[7] T. B. Nguyen, M. A. Pai, and I. A. Hiskens, "Sensitivity approaches for direct computation of critical parameters in a power system," *Int. J. Electr. Power Energy Syst.*, vol. 24, no. 5, pp. 337–343, Jun. 2002.
[8] M. J. Laufenberg and M. A. Pai, "A new approach to dynamic security assessment using trajectory sensitivities," in *Proceedings of the 20th International Conference on Power Industry Computer Applications*, 1997, pp. 272–277.
[9] S. Sharma, S. Pushpak, V. Chinde, and I. Dobson, "Sensitivity of Transient Stability Critical Clearing Time," *IEEE Trans. Power Syst.*, pp. 1–1, 2018.
[10] C. Mishra, R. S. Biswas, A. Pal, and V. A. Centeno, "Critical Clearing Time Sensitivity for Inequality Constrained Systems," *IEEE Trans. Power Syst.*, pp. 1–1, 2019.
[11] V. Venkatasubramanian, H. Schättler, and J. Zaborszky, "Stability Regions for Differential-Algebraic Systems," in *Systems, Models and Feedback: Theory and Applications: Proceedings of a U.S.-Italy Workshop in honor of Professor Antonio Ruberti, Capri, 15–17, June 1992*, A. Isidori and T.-J. Tarn, Eds. Boston, MA: Birkhäuser Boston, 1992, pp. 385–402.
[12] C. Mishra, J. S. Thorp, V. A. Centeno, and A. Pal, "Estimating Relevant Portion of Stability Region using Lyapunov Approach and Sum of Squares," in *2018 IEEE Power Energy Society General Meeting (PESGM)*, 2018, pp. 1–5.
[13] I. A. Hiskens and M. A. Pai, "Trajectory sensitivity analysis of hybrid systems," *IEEE Trans. Circuits Syst. Fundam. Theory Appl.*, vol. 47, no. 2, pp. 204–220, Feb. 2000.
[14] H.-D. Chiang and L. F. Alberto, *Stability regions of nonlinear dynamical systems: theory, estimation, and applications*. Cambridge University Press, 2015.